\documentclass[
reprint,
amsmath,amssymb,
prb,
]{revtex4-2}
\usepackage[utf8]{inputenc}
\usepackage{tikz}
\usepackage[export]{adjustbox}
\usepackage{ulem}
\usepackage{physics}
\usepackage{graphicx,graphics}
\usepackage{dcolumn}
\usepackage{bm}
\usepackage{color}
\usepackage{lipsum}
\usepackage[percent]{overpic}
\usepackage{array}
\usepackage{nicefrac}
\usepackage{multirow}
\usepackage{latexsym,verbatim}
\usepackage{amsmath,amssymb,amsfonts}
\usepackage{xcolor}
\usepackage[breaklinks=true,colorlinks,citecolor=blue,linkcolor=blue,urlcolor=blue]{hyperref}

\makeatletter
\renewcommand{\thefigure}{\arabic{figure}}       
\renewcommand{\fnum@figure}{Figure~\thefigure}   
\makeatother

\begin{document}

\newcommand{\Ts}{T_{\rm s}}
\newcommand{\Tg}{T_{\rm g}}
\newcommand{\Tb}{T_{\rm b}}
\newcommand{\Ps}{P_{\rm s}}
\newcommand{\nep}{{\rm NEP}}
\newcommand{\Qgis}{Q_{\rm GIS}}
\newcommand{\R}{\mathcal{R}}
\newcommand\red[1]{\textcolor{red}{\sout{#1}}}
\newcommand{\blue}{\textcolor{blue}}

\title{Graphene-Insulator-Superconductor junctions as thermoelectric bolometers}
\author{Leonardo Lucchesi}
\affiliation{Dipartimento di Fisica, Università di Pisa, Largo Bruno Pontecorvo 3, 56127 Pisa, PI, Italy}
\affiliation{INFN Sezione di Pisa, Largo Bruno Pontecorvo 3, 56127 Pisa, PI, Italy}
\author{Federico Paolucci}
\affiliation{Dipartimento di Fisica, Università di Pisa, Largo Bruno Pontecorvo 3, 56127 Pisa, PI, Italy}
\affiliation{INFN Sezione di Pisa, Largo Bruno Pontecorvo 3, 56127 Pisa, PI, Italy}

\begin{abstract}
We design a superconducting thermoelectric bolometer made of a Graphene-Insulator-Superconductor tunnel junction. Our detector has the advantage of being passive, as it directly transduces input power to a voltage without the need to modulate an external bias. We characterize the device via numerical simulation of the full nonlinear thermal dynamical model of the junction, considering heating of both sides of the junction . While estimating noise contributions, we found novel expressions due to the temperatures of both sides being different than the bath temperature. Numerical simulations show a Noise Equivalent Power $\nep\sim 4\times 10^{-17}\,{\rm W}/\sqrt{\rm Hz}$ for an input power of $\sim10^{-16}\,{\rm W}$, a response time of $\tau_{th}\sim 200\, {\rm ns}$ and an integration time to obtain a Signal-to-Noise Ratio ${\rm SNR}=1$ of $\tau_{\rm SNR=1}\sim 100\,\mu{\rm s}$ for an input power $\sim 10^{-13}\,{\rm W}$. Therefore, the device show promise for large-array cosmological experiment applications, also considering its advantages for fabrication and heat budget.
\end{abstract}
\maketitle
\section{Introduction}
Research on terahertz detection has grown rapidly in recent years due to its relevance in both fundamental science \cite{P.A.R.Ade2014,Madhavacheril2015} and a wide range of technological applications \cite{Farrah2019,Luukanen2012,Ullom2015}. In this frequency range, thermal detectors such as bolometers become especially important because of the so-called terahertz gap, namely the lack of reliable solid-state devices based on direct photon absorption (see \cite{Rogalski2022} p.55) . For Cosmic Microwave Background (CMB) measurements, present requirements for Noise Equivalent Power (NEP) and saturation power typically lie around $\nep\sim10^{-17}\,{\rm W}/\sqrt{{\rm Hz}}$ and $P_{\rm sat}\sim10\,{\rm pW}$ \cite{Grace2014,Dutcher2024,DeLucia2024}, whereas upcoming experiments aim to introduce stricter requirements, with $\nep\sim3\cdot10^{-20}\,{\rm W}/\sqrt{{\rm Hz}}$ and $P_{\rm sat}\sim0.2\,{\rm fW}$ \cite{Bradford2021}. The Transition-Edge Sensor (TES) remains the benchmark technology for bolometry in astroparticle physics \cite{Irwin1995,DeLucia2024}, with a measured best performance of  $\nep=3\times 10^{-19}\,{\rm W}/\sqrt{\rm Hz}$ \cite{Karasik2011}. However, its operation is complicated by the difficulty of DC readout, by the need for biasing electronics, and the heating associated with it \cite{Heikkilae2018}, while for ground-based experiments the NEP is still limited to$\sim 10^{-17}\,{\rm W}/\sqrt{\rm Hz}$ by environmental factors (see \cite{DeLucia2024} and references within).  \\
An alternative approach is offered by Kinetic Inductance Detectors (KIDs), showing a record $\nep\sim 3\times 10^{-19}\,{\rm W}/\sqrt{\rm Hz} $ \cite{HaileyDunsheath2021}. They have already been deployed—for example, in the OLIMPO experiment, achieving $\nep=4.5\cdot10^{-17}\,{\rm W}/\sqrt{{\rm Hz}}$ at $150,{\rm GHz}$ with an input power of $P=3\,{\rm pW}$ \cite{Paiella2019}. KIDs alleviate the readout challenges and mitigate heating issues typical of TES-based systems. However, their physics is quite complex, and they require costly electronics and wiring for probe signals. They can absorb photons directly by destroying a Cooper pair, reducing the need for antennas, but only above the superconducting gap $\nu>\Delta$ \cite{Wandui2020} (for Al, $\nu>90\,{\rm GHz}$), which poses limitations for CMB applications.\\
Another class of detectors employs Superconducting Tunnel Junctions (STJs). Their working principle relies on the strong temperature dependence of STJ conductivity, and various junction types have been explored, including metal–insulator–superconductor (NIS) devices used in hot-electron bolometers \cite{Golubev2001} and cold-electron bolometers \cite{Kuzmin2004,Kuzmin2019}. Both these designs require voltage or current biasing, with the same problems mentioned for the other technologies.\\
A different approach to detection is represented by passive detectors which directly generate the signal from the input power without the need to modulate a bias. A passive detector concept is the superconducting thermoelectric bolometer \cite{Heikkilae2018}, which measures the thermoelectric current or voltage produced by heating induced by the incident signal. Such a detector could, in principle, overcome several of the issues described above. Among the possible implementations, STJ-based structures are particularly promising, as strong thermoelectric effects arise in Ferromagnet–Insulator–Superconductor (FIS) junctions \cite{Ozaeta2014}. A further mechanism for strong thermoelectricity, accessible in Superconductor–Insulator–Superconductor$^\prime$ (SIS$^\prime$) junctions \cite{Marchegiani2020}, is nonlinear thermoelectricity, obtained by heating the superconductor with the larger $\Delta$. Peculiarly, the induced thermovoltage scales with the superconducting gap $\Delta$ rather than with the temperature $T_{\rm h}$ of the hotter electrode \cite{Marchegiani2020}. This enables sizeable thermovoltages even for weak heating, since a small increase in $T_{\rm h}$ can still drive a signal of order $\sim\Delta$. The main limitation of FIS and SIS$^\prime$ junctions, however, is the requirement of a strong magnetic field within the junction—needed either to activate thermoelectricity (FIS) or to suppress the Josephson current (SIS$^\prime$).\\
In this article, we propose the design of a thermoelectric bolometer made of a Graphene-Insulator-Superconductor (GIS) tunnel junction. There are already several examples of GIS junction-based bolometers in the literature \cite{Vora2012,McKitterick2015,Vischi2020}, but they generate signals from the temperature dependence of resistance measured via biasing and not from thermoelectricity generation.\\
We begin our work by describing the thermal dynamics of the device by using a numerical nonlinear model for thermoelectricity in superconducting junctions introduced in Ref.~\cite{Lucchesi2025}, and then we discuss the noise sources we considered and how we estimated their contributions to the total noise of the device.
Finally, we present the main features and figures of merit of the device, namely the responsivity, the noise equivalent power, and the characteristic times. 
\begin{figure}[bt]
\centering
\includegraphics[width=\columnwidth]{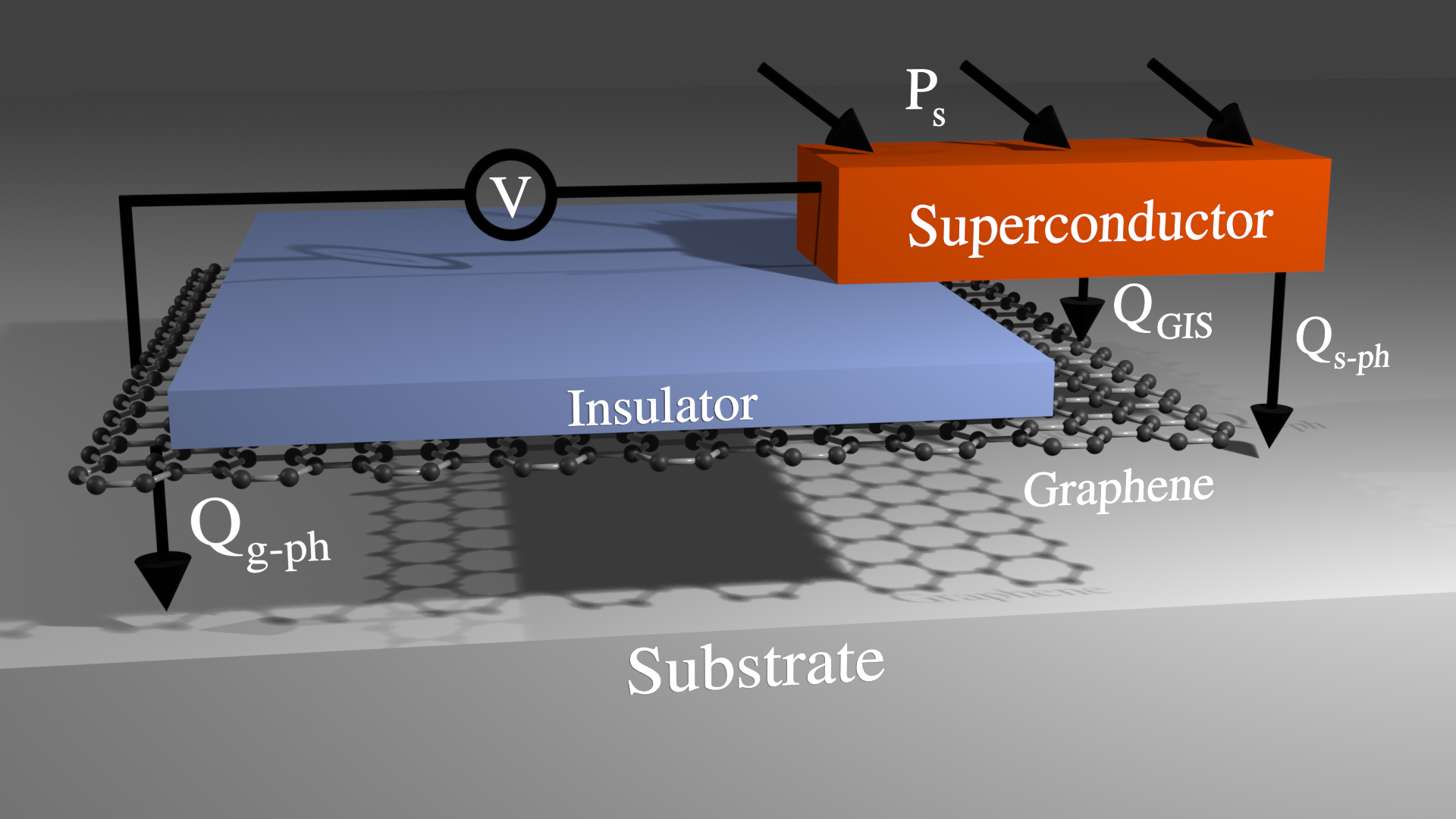}
\caption{Depiction of the thermoelectric GIS bolometer schematizing its thermal dynamics and the measurement setup. 
}
\label{fig2leads}
\end{figure}

\section{Signal model and results}\label{sec:signal}

When one side of a GIS junction heats up, a thermoelectric current starts to flow through it \cite{Bianco2024}. Therefore, we can design a thermal radiation detector by coupling an antenna to one side of the junction and measuring the generated closed-circuit Peltier current $I_0$ or the open-circuit Seebeck potential $V_S$. The thermoelectric current or potential generated by the heat carried by the impinging radiation represents the entire signal of our device. Intuitively, this means that we need the largest possible difference between the superconductor temperature $\Ts$ and the graphene temperature $\Tg$, thus implying that we will need at least one temperature to be much larger than the bath temperature $\Tb$. As shown in Ref.~\cite{Lucchesi2025}, linear transport models fail to describe the thermal behavior of the system in this regime. Following that discussion, we use the nonlinear numerical model defined in Ref.~\cite{Lucchesi2025} to describe the junction behavior and estimate the noise performance of the detector. Here, we only summarize its main points.\\
Current and heat transport across the GIS junction can be described by the usual tunneling formulas \cite{Tinkham,Lucchesi2025}
\begin{multline}\label{eq:iv}
    I(V,\Ts,\Tg)=\frac{1}{eR_T}\int\,dE\; \rho_{\rm s}(E,\Ts)\cdot \\
    \cdot \rho_{\rm g}(E-E_F-eV)[f(E-eV,\Tg)-f(E,\Ts)],
\end{multline}
\begin{multline}\label{eq:qgis}
    Q_{\rm GIS}(V,T_{\rm s},T_{\rm g})=\frac{1}{e^2R_T}\int dE \; (E-eV)\rho_{\rm g}(E-E_F-eV)\cdot\\ \cdot \rho_{\rm s}(E,T_{\rm s}) \left[f(E,T_{\rm s})-f(E-eV,T_{\rm g})\right],
\end{multline}
where $R_T$ is the tunneling resistance, $f(E,T)$ is the Fermi function, $V$ is the electrostatic potential across the junction, $E_F$ is the Fermi energy, $\rho_S$ and $\rho_g$ are the densities of states (DOSs) of the superconductor and graphene defined by Eqs. 2c, 4 and 8 of Ref. \cite{Lucchesi2025}. We consider here doped graphene with electron density $n_0=1\times10^{12}\,{\rm cm}^{-2}$ corresponding to $E_F=98\,{\rm meV}$. The effects of different values of $E_F$ were explored in Ref.~\cite{Bianco2024}. We also choose Al as superconductor, with DOS at the Fermi energy in the metallic state $N_F=2.15\times 10^{47}\,{\rm J}^{-1}{\rm m}^{-3}$ \cite{DeSimoni2018}, and zero-temperature energy gap $\Delta_0=2\times 10^{-4}\,{\rm eV}$. \\
From these expressions, we can obtain $I_0=I(0,\Ts,\Tg)$ and $V_S$ by solving $I(V_S,\Ts,\Tg)=0$ for any $\Ts$ and $\Tg$. In both open and closed-circuit operation, we assume that we have an ideal voltmeter and an ideal ammeter connecting the superconductor and the graphene layers, respectively. Therefore, to compute the response of our device to an input power, we need to describe the thermal dynamics of the junction. Each side of the junction exchanges a power $Q_{\rm GIS}$ with the other side, following Eq.~\ref{eq:qgis}, and dissipates a power $Q_{\rm s-ph}(\Ts)$ (Eq. 6 \cite{Lucchesi2025}) and $Q_{\rm g-ph}(\Tg)$ (Eq. 7a \cite{Lucchesi2025}) through interaction with the phonon gas, which is at the bath temperature $\Tb$ if we assume a negligible Kapitza resistance \cite{Vischi2020}. The physics translates into a system of two coupled nonlinear differential equations for the time-dependent temperatures $T_{\rm s,g}(t)$ that we solve numerically \cite{Lucchesi2025}
\begin{equation}\label{eq:thermdyn}
    \begin{cases}
      \displaystyle C_{\rm s}(T_{\rm s}) \frac{d\Ts}{dt}=-Q_{\rm GIS}(T_{\rm s},T_{\rm g})-Q_{\rm s-ph}(T_{\rm s})+\Ps \\[8pt]
      \displaystyle C_{\rm g}(T_{\rm g}) \frac{dT_{\rm g}}{dt}=Q_{\rm GIS}(T_{\rm s},T_{\rm g})-Q_{\rm g-ph}(T_{\rm g})+P_{\rm g}, 
    \end{cases}
\end{equation}
where $C_{\rm g}(T)=A_{\rm g}\frac{\pi^2}{3}k_B^2 n_{\rm g}(E_F)T$ \cite{Vischi2020} is the thermal capacitance of the graphene layer, with $A_{\rm g}$ the area of the graphene layer, $n_{\rm g}(E_F)$ the density of electrons at the Fermi energy, and $k_B$ the Boltzmann constant 
, and $C_{\rm s}(T)=T dS_{\rm s}/dT$ is the thermal capacitance of the superconductor layer, computed from the quasiparticle entropy $S_{\rm s}(T)$ \cite{Paolucci2023}.
\begin{figure}[tb]
\centering

\begin{tikzpicture}
  \node[anchor=south west, inner sep=0] (img) at (0,0)
    {\includegraphics[width=\columnwidth]{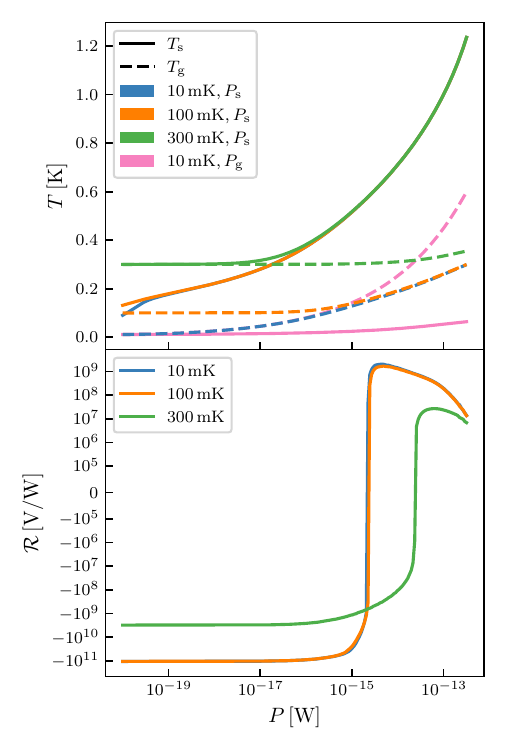}};
  \begin{scope}[x={(img.south east)}, y={(img.north west)}]
    \node[anchor=north west] at (0.02,0.98) {\textbf{a)}};
    \node[anchor=north west] at (0.02,0.52) {\textbf{b)}};
  \end{scope}
\end{tikzpicture}
\caption{Device response to steady input power ${\rm P}$. a) Dependence on ${\rm P}$ of superconductor and graphene temperatures $\Ts,\Tg$ for different $\Tb$ and antenna placement. Temperatures are reasonably smooth in every case, with converging $\Ts$ for larger ${\rm P}$. Connecting the antenna to graphene reduces the thermal gradient. b) Responsivity $\mathcal{R}$ dependence on ${\rm P}$ for different $\Tb$ with antenna on superconductor. $\mathcal{R}$ shows large negative values for small ${\rm P}$, then becomes positive passing from zero. This is due to a saturation effect described in the main text and in Ref.~\cite{Lucchesi2025}.}
\label{fig2}
\end{figure}
$P_{\rm s,g}$ are the external radiant powers impinging on the superconductor and the graphene. We will only explore results with $P_{\rm g}=0$ and $\Ps\neq 0$, because with $\Tg>\Ts$ the absence of nonlinear thermoelectricity reduces the signal \cite{Lucchesi2025}, as also explained below. We will also restrict our analysis to the open-circuit configuration because nonlinear thermoelectricity in this system induces relatively larger $V_S$ than $I_0$ \cite{Lucchesi2025}.\\
For a bolometer, we can assume that the variation of $\Ps$ is slower than the system response time $\tau_{th}$ (and we will later prove that $\tau_{th}$ is reasonably small enough), allowing us to use the steady-state values $\Ts(t\rightarrow \infty)\equiv\Ts$ and $\Tg(t\rightarrow \infty)\equiv\Tg$ to compute all physical quantities. We numerically solve Eq. \ref{eq:thermdyn} for different $\Tb$ to obtain the dependence of $\Ts$ and $\Tg$ on $\Ps$, shown in Fig.~\ref{fig2}a). As we can see from the plot, $\delta T$ is monotonic in $\Ps$, but the dependence is not linear as predicted by linear models \cite{Heikkilae2018}, even for very small $\Ps$ \cite{Lucchesi2025}. We can also notice the effect of the difference in $\Tb$ vanishing for larger $\Ps$, where dissipation processes dominate the thermal dynamics of the system. We choose a specific geometry with $A_{\rm g}=100\,\mu{\rm m}^2$, a superconductor volume of $\mathcal{V}_{\rm s}=0.5\times 10^{-3}\,\mu{\rm m}^3$, and overlap area $A_j=0.01\,\mu{\rm m}^2$ because this geometry allows the formation of the largest thermal gradient $\delta T=\Ts-\Tg$ and gives the best results in terms of signal and noise. Some effects of different geometries are analyzed in Ref.~\cite{Lucchesi2025}. We will use this geometry from now on.\\
The most representative figure of merit for the signal is the responsivity $\mathcal{R}\equiv\delta V_S/\delta \Ps$, defined as the variation in the signal (in open-circuit $\delta V_S$) induced by a small variation in the input power $\delta \Ps$. Since we consider this small variation to happen over a background power $\Ps$, we can compute the system behavior by linearizing Eq.\ref{eq:thermdyn} around the steady-state solutions $\{\Ts(\Ps),\Tg(\Ps)\}$. After a Fourier transform $t\rightarrow \omega$, we can describe the system's linear response $\{\delta \Ts,\delta \Tg\}$ around a point $\{\Ts,\Tg\}$ to power oscillations $\delta \Ps (\omega)$ and $\delta P_{\rm g}(\omega)$ by linearizing Eq.\ref{eq:thermdyn} as
\begin{equation}\label{eq:linthermdyn}
    \begin{cases}
      \displaystyle i\omega C_{\rm s}\delta \Ts=-\left(\frac{\partial Q_{\rm GIS}}{\partial \Ts}+\frac{\partial Q_{\rm s-ph}}{\partial \Ts}\right)\delta\Ts-\frac{\partial Q_{\rm GIS}}{\partial \Tg}\delta\Tg+\delta\Ps \\[8pt]
      \displaystyle i\omega C_{\rm g}\delta \Tg=\left(\frac{\partial Q_{\rm GIS}}{\partial \Tg}-\frac{\partial Q_{\rm g-ph}}{\partial \Tg}\right)\delta\Tg+\frac{\partial Q_{\rm GIS}}{\partial \Ts}\delta\Ts+\delta P_{\rm g}. 
    \end{cases}
\end{equation}
By using the linearized thermal dynamics, we can describe the linear variation of the signal $\delta V_S$ in terms of the linear response functions $\Theta_{gs}(\Ts,\Tg,\omega)\equiv \delta \Tg/\delta \Ts$ and $\Theta_{sp}(\Ts,\Tg,\omega)\equiv \delta \Ts/\delta \Ps$ (expressions in SM \cite{sm}), obtaining
\begin{equation}
    \mathcal{R}=\left(\frac{\partial V_S}{\partial \Ts} + \frac{\partial V_S}{\partial \Tg}\Theta_{gs}\right)\Theta_{sp},
\end{equation}
where we compute the derivatives of $V_S$ from numerically computing $V_S$ on a $(\Ts,\Tg)$ grid \cite{Lucchesi2025}. We show that the dependence of $\mathcal{R}$ on $\Ps$ in Fig.\ref{fig2}b). $\mathcal{R}$ assumes very large negative values $\sim -10^{11}\,{\rm V}/{\rm W}$ for low powers up to $\Ps\sim 1\,{\rm fW}$. For powers larger than $\Ps\sim 1\,{\rm fW}$, $\mathcal{R}$ inverts its sign and reaches relatively large positive values $\mathcal{R}\sim 10^9\,{\rm V}/{\rm W}$. This sign inversion is a nonlinear effect, described in Ref.~\cite{Lucchesi2025}, due to the fact that for larger powers, $V_S(\Ps)$ is not a monotonic function \cite{Lucchesi2025}. Intuitively, for larger $\Ps$, the dominant contribution to changes in $V_S$ is caused by the change in $\Tg$ and not by the change in $\Ts$. This results in an inverted behavior, as an increase in $\Ps$ implies an increase in $\Tg$, reducing the thermal gradient and thus the absolute value of  $V_S$. \\
These zeroes in $\mathcal{R}$ have unfavourable implications in the noise figures of merit, as we will see below. However, the system can be used as a bolometer in both regimes as long as the power is far enough from the inversion point where $\mathcal{R}=0$. 

\section{Noise model and results}
\begin{figure}[tb]
\centering
\begin{tikzpicture}
  \node[anchor=south west, inner sep=0] (img) at (0,0)
    {\includegraphics[width=\columnwidth]{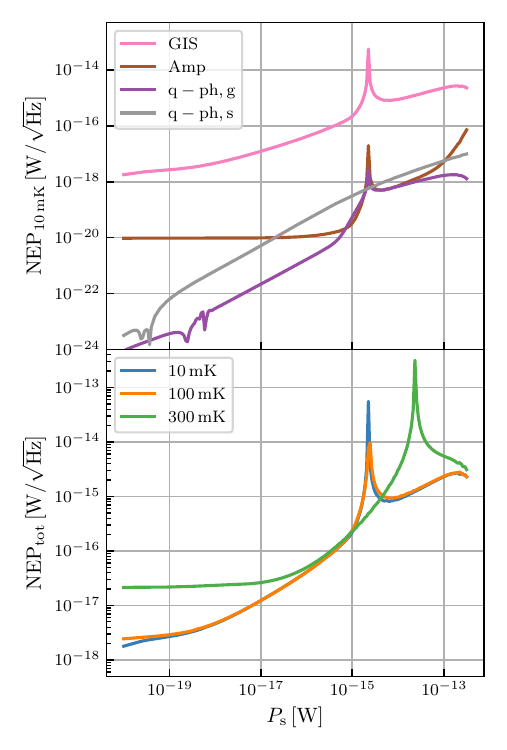}};
  \begin{scope}[x={(img.south east)}, y={(img.north west)}]
    \node[anchor=north west] at (0.02,0.98) {\textbf{a)}};
    \node[anchor=north west] at (0.02,0.52) {\textbf{b)}};
  \end{scope}
\end{tikzpicture}
\caption{Partial and total NEP dependence on $\Ps$. a) NEP contribution at $\Tb=10\,{\rm mK}$ for each noise source. Junction noise dominates by several orders of magnitude because of a large dynamical resistance $R_d$. The peak represents the zero of the responsivity $\mathcal{R}$. b) Comparison between total NEP dependences on $\Ps$ for different $\Tb$. The best reasonable value for the NEP is $4\times 10^{-17}\,{\rm W}/\sqrt{\rm Hz}$ for $\Ps\sim10^{-16}\,{\rm W}$ for $\Tb=100\,{\rm mK}$, but the choice for $\Tb$ could depend on the required input power for the device.}
\label{fig3}
\end{figure}
To assess device properties, modeling noise has the same importance as modeling the signal. While physical noise sources are the same at room temperature and at cryogenic temperature, dominant contributions dramatically change for temperatures lower than the Bloch-Gr{\" u}neisen temperature \cite{Golwala1997,Golubev2001,Giazotto2006,Vischi2020}.  As stated above, we focus on the open-circuit configuration.\\
Noise registered in the measurement comes from charge and energy fluctuations in the quasiparticle gases on both sides of the junction. Charge fluctuations are directly measured via the induced potential fluctuations $\delta V_S$, while the effect of energy fluctuations on steady-state temperatures can be described via Eq. \ref{eq:linthermdyn}, if the variation is slow enough ($\omega \ll 1/\tau_{\rm tun}$, where $\tau_{\rm tun}=e/I\sim 10^{-9}\,{\rm s}$ is the average time between electron tunneling events). \\
We can compare the effects of very different noise contributions by showing their Noise Equivalent Power (NEP), an important bolometric figure of merit \cite{Golubev2001,Heikkilae2018}  defined as the input power $\delta \Ps$ required to generate a signal equal to the average value of the noise $\delta V_S$ with a $1\,{\rm Hz}$ sampling interval.  Following standard literature \cite{Golubev2001,Vischi2020}, we now describe how we estimate the NEP for each dominant noise contribution before computing the total NEP by quadrature sum of all contributions \cite{Mather1982}
\paragraph{Johnson-Nyquist noise}
Any resistive element with resistance $R$ in a circuit induces an open-circuit voltage noise $\langle \delta V^2\rangle=4 k_B R T \delta \omega$ \cite{Nyquist1928}. With our geometry, the graphene layer induces a voltage noise of $\delta V_{\rm g}\equiv\sqrt{\langle\delta V^2\rangle}\sim 5.9\times 10^{-10}\,{\rm V}/\sqrt{\rm Hz}$ at $\Tg=1\,{\rm K}$ if we assume a graphene mobility of $\mu_{\rm g}=10^3\,{\rm cm}^2/{\rm V}\cdot{\rm s}$ and a carrier density $n_0=10^{12}\,{\rm cm}^{-2}$. However, this noise has to be summed incoherently with the noise from the amplifier circuit required to read $V_S$, which we can estimate to be $\delta V_{\rm Amp}\sim 10^{-9}\,{\rm V}/\sqrt{\rm Hz}$ \cite{Paolucci2020}. Therefore, we only keep $\delta V_{\rm Amp}$ for estimation purposes. We can then compute the corresponding NEP by using the definition of responsivity ${\rm NEP}_{\rm Amp}=\delta V_{\rm Amp}/\mathcal{R}$ \cite{Golubev2001}.
\paragraph{Phonon noise}
The interaction between quasiparticles and phonons induces energy fluctuations in a quasiparticle gas \cite{Mather1982,Richards1994,Golwala1997}. These fluctuations happen in both the superconductor and the graphene layer, and they respectively depend on $\Ts$ and $\Tg$. We assume a negligible Kapitza resistance, therefore we consider the phonon gas to be at the thermal bath temperature $\Tb$.\\
For the superconductor layer, we estimate the phonon noise power spectral density by using the equilibrium formula $\delta P_{\rm q-ph,s}^2=4k_B G_{\rm q-ph,s}\Ts^2 \delta \omega$ \cite{Richards1994, Golwala1997, Golubev2001}, with the quasiparticle-phonon thermal conductivity defined as $G_{\rm q-ph,s}=\partial Q_{\rm s-ph}/\partial \Ts$. Even if the quasiparticles and the phonons are not at equilibrium when $\Ts\neq\Tb$, we believe this formula to be a cautious approximation because it corresponds to considering the phonon gas to be at the same temperature as the quasiparticle gas, therefore inducing a larger number of interactions and thus strongly overestimating noise. This noise induces power fluctuations $\delta P_{\rm q-ph,s}$ in the superconductor gas, and therefore it adds to the signal $\Ps$, thus implying ${\rm NEP}_{\rm q-ph,s}=\delta P_{\rm q-ph,s}$.\\
For the graphene layer, we estimate the phonon noise power spectral density by using the Eq. 42 from \cite{Vischi2020}, with $\delta=4$, $\delta P_{\rm q-ph,g}^2=8k_B A_{\rm g} \Sigma_C(\Tg^5+\Tb^5)$, where $\Sigma_C$ is obtained in Eq. 13 of the same article. This noise acts as an oscillating $\delta P_{\rm g}$ in the second line of Eq. \ref{eq:linthermdyn}. This means that it does not generate a $\delta V_S$ with the same mechanism as a $\delta \Ps$ oscillation, and that we cannot equate the NEP to $\delta P_{\rm q-ph,g}$, and we need to compute the response function $\mathcal{R}_{\rm g}\equiv \delta V_S/\delta P_{\rm g}$ and translate the $\delta V_S$ oscillation back to an equivalent input power fluctuation $\delta V_S/\mathcal{R}\equiv\nep$, which is the NEP by definition. After algebraic manipulation of Eq. \ref{eq:linthermdyn}, we obtain
\begin{equation}\label{eq:Rg}
\mathcal{R}_{\rm g}=\left [ \frac{\partial V_S}{\partial \Tg}\left( 1-\frac{\partial Q_{\rm GIS}}{\partial \Tg} \Theta_{gs}\right) - \frac{\partial V_S}{\partial \Ts} \frac{\partial Q_{\rm GIS}}{\partial \Tg} \Theta_{sp} \right ] \mathcal{K}_{\rm g},
\end{equation}
where the expression for $\mathcal{K}_{\rm g}(\Ts,\Tg,\omega)$ can be found in the SM \cite{sm}. We can then compute ${\rm NEP}_{\rm q-ph,g}=(\mathcal{R}_{\rm g}/\mathcal{R})\,\delta P_{\rm g}$.

\paragraph{Junction noise}
The stochasticity of the tunneling process implies that the amount of charge and energy flowing through the junction fluctuates \cite{Golwala1997,Blanter2000}.  
We estimate the contribution to the total NEP by using the fluctuations in current $\langle\delta I^2\rangle$, power $\langle\delta P_j^2\rangle$ and their cross correlation $\langle\delta P_j \delta I\rangle$ from Eqs. 38, 39, 40 of \cite{Vischi2020}, computed from the statistical fluctuations in the number of tunneling quasiparticles by weighing the fluctuations on charge and energies \cite{Golwala1997,Golubev2001}.
\\
Similarly to ${\rm NEP}_{\rm q-ph,g}$, $\delta P_j$ does not simply enter Eq. \ref{eq:linthermdyn} as a $\delta \Ps$ oscillation, but rather it acts as a simultaneous power oscillation on both sides $\delta P_{\rm g}=-\delta \Ps=\delta P_j$ because of energy conservation. We use algebra analogous to Eq. \ref{eq:Rg} to compute the response function $\mathcal{R}_{P_j}\equiv\delta V_S/\delta P_j$, obtaining
\begin{multline}
	\mathcal{R}_{P_j}=- \frac{\partial V_S}{\partial \Ts}\left(\frac{\partial Q_{\rm GIS}}{\partial \Tg} \mathcal{K}_{\rm g}+1\right)\Theta_{sp}+\\
	+\frac{\partial V_S}{\partial \Tg}\left[\mathcal{K}_{\rm g}-\Theta_{gs}\Theta_{sp}\left(\frac{\partial Q_{\rm GIS}}{\partial \Tg}\mathcal{K}_{\rm g}+1\right)\right].
\end{multline}
While we compute the function for the case $P_{\rm g}=0,\Ps\neq 0$, we can find analogous functions for different device configurations. \\
We estimate the open-circuit voltage noise from current fluctuations by using the standard formula $\delta V_{I}=(dI/dV)^{-1}\delta I\equiv R_d \delta I$ \cite{Golwala1997,Golubev2001,Blanter2000}, where we use the derivative of Eq. \ref{eq:iv}. Considering that we can use the same rules for the cross-correlation \cite{Golubev2001},  the final result for the NEP corresponding to junction noise is
\begin{equation}
	{\rm NEP}_j^2=\frac{\mathcal{R}_{P_j}^2\langle\delta P_j^2\rangle + R_d^2\langle\delta I^2\rangle  - 2 \mathcal{R}_{P_j}R_d\langle\delta P_j \delta I\rangle}{\mathcal{R}^2},
\end{equation}
\begin{figure*}[t]
\centering
\begin{tikzpicture}
  \node[anchor=south west, inner sep=0] (img) at (0,0)
    {\includegraphics[width=\textwidth]{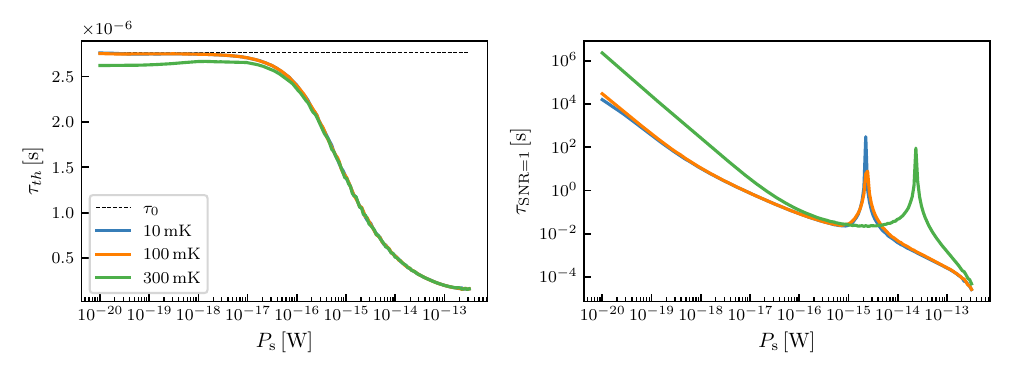}};
  \begin{scope}[x={(img.south east)}, y={(img.north west)}]
    \node[anchor=north west] at (0.02,0.98) {\textbf{a)}};
    \node[anchor=north west] at (0.52,0.98) {\textbf{b)}};
  \end{scope}
\end{tikzpicture}
\caption{Characteristic times dependence on $\Ps$. a) Thermal response time $\tau_{th}$ for different $\Tb$. The curves for $10$ and $100$ mK are almost identical. For small $\Ps$ and $\Tb$,  $\tau_{th}(\Ps)\simeq\tau_0$ shows almost no temperature dependence (see main text). $\tau_{th}$ drops for larger $\Ps$ because of increased energy dissipation in the superconductor. b) Integration time necessary to get a signal-to-noise-ratio equal to 1 $\tau_{\rm SNR=1}$. While our bolometer shows very large integration times $\sim 10^3\,{\rm s}$ for smaller powers, the integration time improves with $\Ps$, dropping down to $\sim 1\,{\rm ms}$ for $\Ps\sim 10^{-13}\,{\rm W}$ and $\Tb=100\,{\rm mK}$. Integration times are computed for signal variations corresponding to the entire signal.}
\label{fig4}
\end{figure*}
where the relative sign between the correlation and the squared noise contributions comes from the tunneling process properties \cite{Golwala1997}.\\
In Fig. \ref{fig3}a, we can see the results for every NEP contribution computed for our device setup with $\Tb=10\,{\rm mK}$. We can see that the junction contribution overwhelmingly dominates over every other noise source. This is mainly due to the large contribution of the $\langle \delta I^2 \rangle$ part, which is multiplied by a large dynamic resistance $R_{d}\sim 1\, {\rm G}\Omega$. This large $R_{d}$ comes from the junction still being away from the Ohmic part of the IV curve at $V=V_S$ for our values of the Seebeck potential. The sharp peaks present in each contribution except ${\rm NEP}_{q-ph,s}$ correspond with the $\mathcal{R}=0$ point shown in Fig. \ref{fig2}b. \\
In Fig. \ref{fig3}b, we represent the total NEP dependence on input power $\Ps$ for different values of $\Tb$.  This is the most important figure of merit for a bolometer, and for our setup its most useful value is ${\rm NEP}_{100\,{\rm mK}}\sim 4\times 10^{-17}\,{\rm W}/\sqrt{\rm Hz}$ estimated for $\Ps\sim10^{-16}\,{\rm W}$ and $\Tb=100\,{\rm mK}$, since it shows similar behavior to $10$ mK. 
Another important feature is that, for intermediate powers $10^{-16}\,{\rm W}\lesssim\Ps\lesssim 10^{-14}\, {\rm W}$, the larger bath temperature $\Tb=300\,{\rm mK}$ shows a performance similar to the smaller $\Tb$, useful for further relaxing requirements on the cryogenic apparatus. 
\section{Characteristic times}
Characteristic times represent another important consideration for a bolometer design. The two most important timescales to represent the behavior of the bolometer are the response time and the integration time needed to obtain a useful measurement.\\
Since electrical response times are much faster, we estimate the bolometer response time via the thermal response time $\tau_{th}$, the time needed by the system to adjust its temperature after a change in the power input. We imagine a response to a small variation $\delta P$ over a larger power input $\Ps$, that makes the system reach a new steady state exponentially via Eq. \ref{eq:thermdyn}, with a characteristic time
\begin{equation}
\tau_{th}=\frac{C_{\rm s}(\Ts)}{G_{\rm GIS}(\Ts,\Tg)+G_{\rm q-ph,s}(\Ts)},
\end{equation}
where $G_i=\partial Q_i/\partial \Ts$. This standard approximation \cite{Mather1982,Richards1994} is reasonable for estimation purposes, and it gives a correct physical picture for experiments where the signal variations are rather slow, such as the sky scans for CMB survey. We represent the dependence of $\tau_{th}$ on $\Ps$ in Fig. \ref{fig4}a. For smaller $\Ps<10^{-17}\,{\rm W}$, $\tau_{th}$ is constant for $\Tb=10,100\,{\rm mK}$, then it drops down by an order of magnitude from $\Ps\sim10^{-17}\,{\rm W}$ to $\Ps\sim10^{-13}\,{\rm W}$.  For $\Tb=300\,{\rm mK}$, $\tau_{th}$ is almost constant. For $\Tb=300\,{\rm mK}$, $\tau_{th}$ shows the same behavior as $\Tb=100\,{\rm mK}$, except for a different limit value at lower $\Ps$, due to non-negligible quasiparticle dissipation in the superconductor (see Fig.~4 in Ref.~\cite{Lucchesi2025}). We proved in the SM \cite{sm} that, for small $\Ps$, $\tau_{th}$ converges to a fixed value $\tau_0=\mathcal{V}_{\rm s}N_F e^2 R_T\simeq 2.76\,\mu{\rm s}$, as long as $G_{\rm q-ph,s}\ll G_{\rm GIS}$ $\Tg\ll\Ts$ and $\Ts\ll \Delta_0/k_B$. In Fig. 4 of \cite{Lucchesi2025}, we can see that $Q_{s-ph,s}$ is very small for $\Ts\lesssim 0.3\,{\rm K}$, and thus $G_{\rm q-ph,s}$ is negligible except for $\Tb=300\,{\rm mK}$. We can see that $\tau_{th}$ reaches $\sim 200\, {\rm ns}$ for $\Ps\sim 10^{-13}\,{\rm W}$. The effect of $G_{\rm q-ph,s}$ also inverts the dependence of $\tau_{th}$ on $\Ps$, making it drop, while $G_j$ alone would make it rise.\\
We define the integration time 
\begin{equation}
\tau_{\rm SNR=1}=\frac{1}{2}\left(\frac{\rm NEP_{tot}}{\Ps}\right)^2
\end{equation}
as the time that a signal of fixed power $\Ps$ needs to be integrated to have a signal-to-noise ratio ${\rm SNR}=1$. Integrating for a longer time reduces the frequency interval and thus reduces the absolute value of the noise power. In Fig. \ref{fig4}b, we show the dependence of $\tau_{\rm SNR=1}$ on $\Ps$. Interestingly, $\tau_{\rm SNR=1}$ becomes smaller with $\Ps$, down to $\sim 100\,\mu{\rm s}$ for $\Ps\sim 10^{-13}\,{\rm W}$. The $\mathcal{R}=0$ peaks are also present in this plot, due to the NEP in the numerator.
\section{Conclusions}
Throughout this article, we have proposed and numerically characterised a thermoelectric bolometer made of a Graphene-Insulator-Superconductor junction. After assessing that we have a better device performance by mounting the antenna on the superconductor side, we discovered that the device shows a large responsivity $\mathcal{R}$ up to $|\mathcal{R}|\sim 10^{11}\,{\rm V}/{\rm W}$ for $\Ps\lesssim 1\, {\rm fW}$ and $\Tb=100\,{\rm mK}$. Furthermore, our device design shows a $\nep\sim 4\times 10^{-17}\,{\rm W}/\sqrt{\rm Hz}$ for $\Ps\sim10^{-16}\,{\rm W}$ and $\Tb=100\,{\rm mK}$, and a $\nep\sim 2\times 10^{-15}\,{\rm W}/\sqrt{\rm Hz}$ for $\Ps\sim10^{-13}\,{\rm W}$. Even if our NEP is larger than the measured $\sim 10^{-17}\,{\rm W}/\sqrt{\rm Hz}$ found in the literature \cite{DeLucia2024} and the requirement for future experiments, it is a passive detector design with performance almost comparable with active detectors. In the preparation of a large array experiment, a smaller impact on the heat budget and cheaper fabrication could be more useful than a smaller NEP, which can, in principle, be compensated by longer integration times. Since this is a preliminary study, further optimization could also improve the noise performance of our device. Our detector shows response times down to $\tau_{th}\sim 200\, {\rm ns}$ for $\Ps\sim 10^{-13}\,{\rm W}$ , and integration times down to $\tau_{\rm SNR=1}\sim 100\,\mu{\rm s}$ for $\Ps\sim 10^{-13}\,{\rm W}$. These values are compatible with investigations of signals arising from experiments focused on interesting physical phenomena, such as the polarization of CMB \cite{Grace2014}. Thus, this technology might be implemented in experiments where the total power budget of the apparatus is limited, such as balloon or satellite-based cosmology experiments.

\begin{acknowledgments}
The authors wish to thank T.T. Heikkil{\"a} for fruitful discussion. The  Italian Ministry of University and Research partially funded the work of L.L. and F.P. under the call PRIN2022 (Financed by the European Union – Next Generation EU) project EQUATE (Grant No. 2022Z7RHRS) and under the call FIS2 project QuLEAP (Grant No. FIS2023-00227). F.P. awknowledges the CSN V of INFN under the technology innovation grant STEEP for partial financial support.
\end{acknowledgments}
\vspace{0.5\baselineskip}
\appendix
\section{Response functions and linearization of thermal dynamics}
To obtain the response functions to small oscillations in input power, we start from the linearized version of the thermal dynamics equations
\begin{equation}\label{eq:linthermdyn}
    \begin{cases}
      \displaystyle i\omega C_{\rm s}\delta \Ts=-\left(\frac{\partial Q_{\rm GIS}}{\partial \Ts}+\frac{\partial Q_{\rm s-ph}}{\partial \Ts}\right)\delta\Ts-\frac{\partial Q_{\rm GIS}}{\partial \Tg}\delta\Tg+\delta\Ps \\[8pt]
      \displaystyle i\omega C_{\rm g}\delta \Tg=\left(\frac{\partial Q_{\rm GIS}}{\partial \Tg}-\frac{\partial Q_{\rm g-ph}}{\partial \Tg}\right)\delta\Tg+\frac{\partial Q_{\rm GIS}}{\partial \Ts}\delta\Ts+\delta P_{\rm g}. 
    \end{cases}
\end{equation}
If we want to understand the response to a variation in $\delta\Ps$, we algebraically solve the system first for $\delta \Tg$, obtaining
\begin{equation}
\delta \Tg = \Theta_{gs}\delta \Ts,
\end{equation}
and then we solve for $\delta \Ts$, obtaining
\begin{equation}
\delta \Ts =\Theta_{sp} \delta \Ps,
\end{equation}
with
\begin{widetext}
\begin{subequations}\label{eqtheta}
	\begin{align}
        \Theta_{gs}(\Ts,\Tg,\omega)\equiv \frac{\delta \Tg}{\delta \Ts}&=\dfrac{\dfrac{\partial Q_{\rm GIS}}{\partial \Ts}}{i\omega C_{\rm g}- \dfrac{\partial Q_{\rm GIS} }{\partial \Tg} + \dfrac{dQ_{\rm g-ph}}{d\Tg}}\\
		\Theta_{sp}(\Ts,\Tg,\omega)\equiv \frac{\delta \Ts}{\delta \Ps}&=\dfrac{1}{i\omega C_{\rm s}- \dfrac{\partial Q_{\rm GIS} }{\partial \Tg} \Theta_{gs}(\Ts,\Tg,\omega) - \dfrac{\partial Q_{\rm GIS} }{\partial \Ts} + \dfrac{dQ_{\rm s-ph}}{d\Ts}}.
	\end{align}
\end{subequations}
\end{widetext}
The same set of equations could be solved differently to obtain the response to a small $\delta P_{\rm g}$ with a fixed $\Ps$. In that case, we still solve first for $\delta \Tg$, but this time we obtain
\begin{equation}
\delta \Tg = \Theta_{gs}\delta \Ts + \mathcal{K}_{\rm g} \delta P_{\rm g},
\end{equation}
with
\begin{equation}
\mathcal{K}_{\rm g}(\Ts,\Tg,\omega)=\dfrac{1}{ i\omega C_{\rm g}-\dfrac{\partial \Qgis}{\partial \Tg}+\dfrac{d Q_{\rm g-ph}}{d \Tg}}.
\end{equation}
The second part is the same as before, and it allows us to find the response functions to power fluctuations in the graphene quasiparticle gas $\mathcal{R}_{\rm g}$ and to the tunneling-induced power fluctuations in the junction $\mathcal{R}_{P_j}$, as described in the main text. 
\section{Demonstration of constant thermal response time for small input powers}
In Fig.4b) of the main text, we can see that for low input powers $\Ps$ and low bath temperatures $\Tb$ the response time is almost constant. The thermal response time of a bolometer is usually defined as \cite{Mather1982}
\begin{equation}
\tau_{th}=\frac{C_{th}}{G_{th}},
\end{equation}
where $C_{th}$ is the thermal capacitance of the active region and $G_{th}$ is the thermal conductance between the charge carriers in the active region and the thermal bath. In the linear regime, $C_{th}$ and $G_{th}$ are computed at the bath temperature, therefore they are constant for every input power. However, as shown in \cite{Lucchesi2025}, the linear regime breaks for power much smaller than the $\Ps\sim 10^{-17}\,{\rm W}$ where we still observe an almost constant $\tau_{th}$. Here, we prove that the constant value of $\tau_{th}$ holds even in the nonlinear regime, as long as $\Ts$ and $\Tg$ are small enough.\\
When $\Ps$ is small, the bottleneck for heat conduction is the junction $G_{\rm GIS}$, as phonon-induced dissipation in the superconductor is negligible and phonon-induced dissipation in graphene is strong enough to keep $\Tg\sim \Tb$. Therefore, the thermal capacitance of the superconductor acts as the thermal capacitance of the entire system, and the Seebeck potential is negligible $V_S\simeq0$. From the definitions we obtain
\begin{widetext}
\begin{align}
C_{\rm s}(\Ts)=\Ts \frac{dS_{\rm s}}{d\Ts}&=-2\mathcal{V}_{\rm s} N_F k_B \Ts\frac{d}{d\Ts} \int_{-\infty}^{+\infty} dE\; \rho_{\rm s}(E) f(E,\Ts)\log{f}(E,\Ts) \label{eq:cs}\\
G_{\rm GIS}(\Ts,\Tg)=\frac{\partial Q_{\rm GIS}}{\partial \Ts}&=\frac{1}{e^2R_t}\frac{\partial }{\partial \Ts}\int dE \; E\,\rho_{\rm g}(E-E_F)\rho_{\rm s}(E) \left[f(E,T_{\rm g})-f(E,T_{\rm s})\right], \label{eq:gth}
\end{align}
\end{widetext}
where $R_T$ is the tunneling resistance, $f(E,T)$ is the Fermi function, , $\rho_S$ and $\rho_g$ are the densities of states (DOSs) of the superconductor and graphene defined by Eqs. 2c, 4 and 8 of Ref. \cite{Lucchesi2025}, $N_F=2.15\times 10^{47}\,{\rm J}^{-1}{\rm m}^{-3}$ is the superconductor DOS at the Fermi energy in the metallic state, and $\mathcal{V}_{\rm s}$ is the supeconductor volume.
Since we consider doped graphene with electron density $n_0=1\times10^{12}\,{\rm cm}^{-2}$, we can assume the relative graphene DOS $\rho_{\rm g}(E-E_F)=1$ to be approximately constant over the relevant energy interval and we consider $\Ts$ to be low enough to not impact on the superconductor DOS $\rho_{\rm s}$.\\
If we focus on the $\Ts$ dependent part of $C_{\rm s}(\Ts)$, we can see that Eq. \ref{eq:cs} can be rewritten as
\begin{align}\label{eq:cs2}
	&\Ts\frac{d}{d\Ts} \int dE\; \rho_{\rm s}(E) f(E,\Ts)\log{f}(E,\Ts)= \nonumber\\
	&\Ts \int dE\; \frac{\partial \rho_{\rm s}}{\partial \Ts} f\log{f} +\rho_{\rm s} \frac{\partial f}{\partial \Ts}\log{f} +\rho_{\rm s}\frac{\partial f}{\partial \Ts}. 
\end{align}
The last term in Eq. \ref{eq:cs2} is zero because $\rho_{\rm s}(E)$ is even in $E$ and $\partial f/\partial \Ts$ is odd in $E$. In the $\Ts\rightarrow 0$ limit ($\Ts\ll \Delta_0/k_B$), we can recast the second term as
\begin{equation}
\lim_{\Ts\rightarrow 0}\,\ln{f}=\begin{cases} -\frac{E}{k_B\Ts}\quad &E>0\\
0\quad &E<0
\end{cases}
\end{equation}
obtaining
\begin{equation}
\lim_{\Ts\rightarrow 0}\, C_{\rm s}(\Ts)= 2\mathcal{V}_{\rm s} N_F \int_{0}^{+\infty} dE\; E\, \left(\frac{\partial \rho_{\rm s}}{\partial \Ts} f +\rho_{\rm s} \frac{\partial f}{\partial \Ts}\right).
\end{equation}
We now focus on the $\Ts$ dependent part of $G_{\rm GIS}$ in Eq. \ref{eq:gth}, 
\begin{align}
	&\frac{\partial }{\partial \Ts}\int_{-\infty}^{+\infty} dE \; E\,\rho_{\rm s}(E) \left[f(E,T_{\rm s})-f(E,T_{\rm g})\right]=\\
	&\int_{-\infty}^{+\infty} dE \; E\,\left[\frac{\partial \rho_{\rm s}}{\partial \Ts} \left(f(E,T_{\rm s})-f(E,T_{\rm g})\right)+\rho_{\rm s}\frac{\partial f}{\partial \Ts}\right].
\end{align}
The second term is even in $E$, and its integral between $-\infty$ and $+\infty$ is just two times its integral between $0$ and $+\infty$. 
If $\Tg\ll \Ts$,
\begin{equation}
\lim_{\Tg\ll\Ts}\, f(E,T_{\rm s})-f(E,T_{\rm g})=\begin{cases}
	f(E,T_{\rm s})\quad &E>0\\
	f(E,T_{\rm s})-1\quad&E<0.
\end{cases}
\end{equation}
Then, we can recast the negative energy part of the integral by using the variable transformation $E\rightarrow-E$ and the fact that $\partial \rho_{\rm s}/\partial\Ts$ is an even function of $E$
\begin{equation}
\int_{-\infty}^0 dE\; E \frac{\partial \rho_{\rm s}}{\partial\Ts}\left(f(E,T_{\rm s})-1\right)=\int_0^{+\infty} dE\; E \frac{\partial \rho_{\rm s}}{\partial\Ts}f(E,T_{\rm s}),
\end{equation}
which is equal to the positive energy part. Therefore
\begin{equation}
\lim_{\Tg\ll\Ts}\, G_{j}=\frac{2}{e^2 R_T} \int_{0}^{+\infty} dE\; E\, \left(\frac{\partial \rho_{\rm s}}{\partial \Ts} f +\rho_{\rm s} \frac{\partial f}{\partial \Ts}\right).
\end{equation}
The $\Ts$ dependent parts of $C_{\rm s}$ and $G_j$ are equal, so if $\Ts\rightarrow 0$ and $\Tg\ll \Ts$, $\tau_{th}$ remains constant at the limiting value
\begin{equation}
\lim_{\Tg\ll\Ts\rightarrow0} \frac{C_{\rm s}}{G_j}=\mathcal{V}_{\rm s} N_F e^2 R_T\simeq2.76\,\mu{\rm s}.
\end{equation}
In our setup, $\Ts$ and $\Tg$ follow the conditions even for $\Ps>10^{-17}\,{\rm W}$, but the thermal conductivity induced by quasiparticle-phonon dissipation in the superconductor for $\Ps>10^{-17}\,{\rm W}$ is not negligible anymore, inducing the change in $\tau_{th}$ described in the main text.

\end{document}